\documentclass[runningheads]{llncs}
\makeatletter
\newcommand{\@chapapp}{\relax}%
\makeatother
\usepackage[title,toc,titletoc,header]{appendix}
\usepackage{graphicx}
\usepackage{float}
\usepackage{hyperref}

\begin{document}

\title{Early Detection of In-Memory Malicious Activity based on Run-time Environmental Features}

\titlerunning{Early Detection of In-Memory Malicious Activity}

\author{Dorel Yaffe \and Danny Hendler}

\authorrunning{D. Yaffe and D. Hendler}

\institute{ Department of Computer Science, Ben-Gurion University, Beer-Sheva, Israel
\email{yaffed@post.bgu.ac.il, hendlerd@bgu.ac.il}}

\maketitle

\begin{abstract}
In recent years malware has become increasingly sophisticated and difficult to detect prior to exploitation \cite{erdHodi2020exploitation}. While there are plenty of approaches to malware detection, they all have shortcomings when it comes to identifying malware correctly prior to exploitation. The trade-off is usually between false positives, causing overhead, preventing normal usage and the risk of letting the malware execute and cause damage to the target \cite{marculet2019methods}\cite{huajmuacsan2017dynamic}\cite{lungana2018false}.\\
We present a novel end-to-end solution for in-memory malicious activity detection done prior to exploitation by leveraging machine learning capabilities based on data from unique run-time logs, which are carefully curated in order to detect malicious activity in the memory of protected processes. This solution achieves reduced overhead and false positives as well as deployment simplicity.\\
We implemented our solution for Windows-based systems, employing multi disciplinary knowledge from malware research, machine learning, and operating system internals. Our experimental evaluation yielded promising results. As we expect future sophisticated malware may try to bypass it, we also discuss how our solution can be extended to thwart such bypassing attempts.

\keywords{Malware Detection \and Machine Learning \and Malicious Activity \and In-memory Attacks \and Log Analysis \and Moving Target Defense \and Forensics \and Run-time Logs}
\end{abstract}

\section{Introduction}
\label{SectionIntroduction}
In recent years malware has become increasingly sophisticated and difficult to detect prior to exploitation\cite{erdHodi2020exploitation}, and there are different approaches that attempt to detect and protect against it prior to exploitation of the target, see e.g. \cite{aslan2020comprehensive}\cite{chakkaravarthy2019survey}\cite{or2019dynamic}. Solutions include Endpoint Protection Platforms (EPP) such as legacy or next-generation anti-viruses, Endpoint Detection and Response (EDR), Managed Detection and Response (MDR), Cross-layered Detection and Response (XDR) and many more \cite{epp_understanding}. These solutions usually rely on either a constantly updated database of signatures for detecting malware variants that have already been seen in the wild, a predefined set of rules or heuristics for identifying malicious activity by actions taken by a program, leveraging more advanced machine learning techniques like behavioral analysis (to deduce if an entity or set of actions is malicious or not), or any hybrid combination of these approaches \cite{irshad2018effective}\cite{naz2019review}\cite{souri2018state}.
\\\\
All these approaches have shortcomings when it comes to identifying malware correctly prior to exploitation. In some cases, if the resource defined as a potential target is very critical to the organization, e.g. a global authentication service running on a secure server, the owners will have a great incentive to protect it, but for most protection mechanisms this is likely to cause an increased number of false positives on accesses of the resource, an adverse impact on performance, or even hurting the functionality provided by the resource. All these potential problems often deter organizations from hardening the protection of critical resource in comparison with the rest of the system \cite{huajmuacsan2018performance}\cite{malware_issues_whitepaper}.
\\\\
The end result is a similar protection level for all of the resources in the system which is often insufficient for protecting against very sophisticated malware, especially in-memory ones \cite{patten2017evolution}\cite{kumar2020emerging}.
\\\\
In this work, we describe a security mechanism that ensures, with high probability, that detection occurs in real time prior to any malicious action that impacts the attacked system.
\\\\
Our solution was implemented by training a malware detection machine learning model on a large data-set of run-time logs (provided by \href{https://www.morphisec.com}{Morphisec Ltd}). These logs were collected from real-world machines during both standard workloads and real malware attacks. The logs contain a set of environmental and process-related variables we efficiently extract during run-time that were identified as important for the detection of in-memory attacks.\\
Using the deployed detector, our system is able to query a minimal run-time log from a critical process to infer if an in-memory malicious activity is about to take place and perform a mitigation action before any harmful operation is performed.
\\\\
Numerous malware detection approaches exist \cite{nigammalware}\cite{mirza2018cloudintell}\cite{niveditha2020detect}\cite{birman2020cost}\cite{erdHodi2020exploitation}\cite{aslan2020comprehensive}\cite{chakkaravarthy2019survey}\cite{or2019dynamic}\cite{irshad2018effective}\cite{naz2019review}\cite{souri2018state}. Malware detection using ML models often analyzes the contents of files (e.g. executable files) or other data sources for extracting features to train the model \cite{sihwail2018survey}\cite{sethi2018novel}\cite{sethi2019novel}\cite{novelmalware}\cite{irshad2019feature}\cite{ijaz2019static}, or transforms the problem to another domain, such as image recognition \cite{zhang2016irmd}\cite{venkatraman2019hybrid}. Unlike these approaches, our solution uses run time environment logs in a very efficient and seamless manner (in comparison to memory dump analysis or sandbox/VM approach \cite{tien2017memory}\cite{jamalpur2018dynamic}\cite{wang2017malware}\cite{walker2019cuckoo}). In comparison to other solutions that monitor processes in real-time, our solution is non-intrusive and does not hurt system performance \cite{marculet2019methods}\cite{huajmuacsan2017dynamic}\cite{lungana2018false}\cite{huajmuacsan2018performance}\cite{malware_issues_whitepaper}. It also reduces false positives to a minimum based on a rating system, as a threshold is set depending on how early the detection is required, providing a trade-off between the length of the early warning interval and the rate of false positives.
\\\\
The rest of this paper is organized as follows: In section \ref{SectionDataset} we describe our data set. In section \ref{SectionMethod} we showcase the high level pipeline for our model creation, features selection and describe the rational behind it. In section \ref{SectionSolution}, we describe the high-level architecture of our detector. We report on our experimental evaluation in section \ref{SectionExperimentalEvaluation}. Finally, we discuss our conclusions and directions for future work in section \ref{SectionConclusionFuture}.

\section{Dataset}
\label{SectionDataset}
The dataset used in this research is composed of logs containing environmental information regarding a specific process, that are collected live at runtime.\\
The environmental information collected consists of both static data (that could have been collected without the execution of the process) and dynamic data collected from the virtual memory of the process itself.\\\\
The dataset spans logs from the entire year of 2019 (that is, between January 1, 2019 and December 31, 2019). These logs are from real world scenarios, real world incidents in actual production machines with real users using them on a daily basis to do their work.\\\\
The logs were collected using a proprietary agent program (provided by \href{https://www.morphisec.com}{Morphisec Ltd}). This proprietary software leverages deep integration in the operating system’s kernel in order to block malicious activity during runtime using various  memory-morphing techniques.
The logs are not part of the defense mechanism, we made use of these logs to further understand and analyse specific points in the life cycle of processes.\\\\
These logs are taken \textit{prior to exploitation}, which means that any malicious activity has not yet taken place in the target machine, however, the malicious entity was indeed running.\\\\
There are two types of logs collected in the dataset:
\begin{enumerate}
    \item \textbf{Malicious activity}: Environmental runtime logs from incidents where the following actions taken by the malicious entity were indeed malicious (e.g. a log from the starting point of a ransomware malware execution).
    \item \textbf{Benign activity}: Environmental runtime logs from incidents where the application/process/service are benign and the following actions were non-malicious (e.g. opening a malware-free document file via Microsoft Word).
\end{enumerate}
Using the proprietary deterministic solution to mark each log, means that all of the records in the dataset are already \textbf{labeled} as \textit{malicious} or \textit{benign}. These labels were reviewed regularly by malware analysts to further reduce the probability of false positives/negatives.

\subsection{Data Analysis}
Each log is unique, there are no two logs which are exactly the same.\\
Any identifiable information in the logs, such as usernames, machine names, domain names, etc. was anonymized in order to preserve privacy.\\
In Table \ref{tbl:dataset_labels} and Table \ref{tbl:dataset_classification} we present some statistics about the dataset.
\begin{table}[H]
            \centering
            \caption{Dataset: Labels}
            \label{tbl:dataset_labels}
            \begin{tabular}{ |c|c| } 
                \hline
                Label & Count \\
                 \hline
                 Malicious	& 1,637,645 \\
                 Benign	& 827,938 \\
                 Total	& 2,465,583 \\
                \hline
            \end{tabular}
%        \end{table}
            \vspace{+1em}
%\begin{table}[H]
%            \centering
            \caption{Dataset: Classification by type}
            \label{tbl:dataset_classification}
            
            \begin{tabular}{ |c|c| } 
                \hline
                Classification by malware & Count \\
                 \hline
                 Classified	& 1,043,520 \\
                 Unclassified	& 594,125 \\
                 Total (malicious) & 1,637,645 \\
                \hline
            \end{tabular}
        \end{table}
As can be seen from Table \ref{tbl:dataset_labels}, there is almost a 2:1 ratio between the malicious samples and the benign ones. As presented by Table \ref{tbl:dataset_classification}, almost two thirds of the malicious samples are also classified by type (e.g. Trojan).
\\\\
In Table \ref{tbl:dataset_unique} we present data showing that our data set is heterogeneous in terms of malware and attack types, operating systems, machines, users etc., thus the data is unlikely to cause over-fitting to any one of these characteristics.
\begin{table}[H]
            \centering
            \caption{Dataset: Unique values}
            \label{tbl:dataset_unique}
            
            \begin{tabular}{ |c|c| } 
                \hline
                Parameter & Count \\
                 \hline
                 Unique Malware Types 	& +15 \\
                 Unique Attacks 	& +100 \\
                 Unique Operating Systems	& +10 \\
                 Unique Machines & +1M \\
                 Unique Users & +1M \\
                 Unique Domains & +100K \\
                 Unique IP Addresses & +1M \\
                 Unique Application/Process Names & +1000 \\
                 Unique Thread Count & +20 \\
                 Unique Loaded Modules Count & +500K \\
                \hline
            \end{tabular}
        \end{table}
There are \textit{20} malware categories in the dataset as classified by YARA rules. These categories are detailed by Table \ref{tbl:dataset_categories}.

\begin{table}[H]
            \centering
            \caption{Dataset: Malware categories}
            \label{tbl:dataset_categories}
            \begin{tabular}{ |c c c c| } 
                \hline
                 Malware Categories & & & \\
                 \hline
                 Adware & Exploit Kit & Keylogger & Shellcode \\
                 Fileless & Vulnerability & Banking Trojan & Info Stealer\\
                 Rootkit & Code Injection & Miner & Backdoor \\
                 Spyware & Supply Chain & Ransomware & Trojan \\
                 Virus & Hacking Tool & Worm & Remote Code Execution \\
                \hline
            \end{tabular}
        \end{table}

\subsection{Log Structure}
\label{sec:log_struct}
The records in the dataset are basically documents in JSON format, resembling runtime logs containing a predetermined set of fields holding environmental data about a certain process under observation. The log files contain a lot of textual information. Nevertheless, their size is typically quite small, peaking at 500KB.\\\\
There are dozens of different fields and attributes as part of the log specification (which is proprietary), however, we chose only the most relevant fields in the logs, detailed in \ref{ChapterAppendixC}, which are also available to be collected during runtime using open source tools and/or low level coding.

\section{Pre-Processing and Model Generation}
\label{SectionMethod}
The essence of the solution lays in the machine learning model trained on a big data set of real-world logs representing malicious activity in various execution stages and from various types of malware, as well as benign data from standard applications.

\subsection{High-level flow}
\begin{figure}[H]
        	\begin{center}
        		\includegraphics[width=0.7\columnwidth]{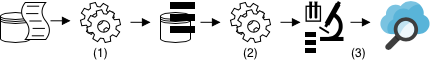}
        	\end{center}
        	\caption{Pre-Processing and Model Generation: Model training process}
        	\label{figure:ml_model}
\end{figure}

The high level flow of the model training procedure is as follows (see Figure \ref{figure:ml_model}):
\begin{enumerate}
    \item Given a data set of logs in the correct format, extract the textual features required and convert them from textual representations to numeric vectors using a standard Natural Language Processing algorithm, do this for every value (i.e. a full word) in the logs.
    \item Given a data set of logs represented as numeric vectors (that is, the output from the previous stage), convert each log file containing these numeric representations to a single numeric vector, using a standard calculation from the world of document classification.
    \item Given a set of numeric vectors to be used as a training set with proper labels (either malware or benign), train a prediction model in a supervised manner, using a standard Machine Learning ensemble approach from the world of classification.
\end{enumerate}

We describe the training process in more detail in the following sub-sections.

\subsection{Data Preparation}
In this stage, we did the following:
\begin{enumerate}
    \item Cleaned the data where it was needed using Python scripting over the JSON documents:
    \begin{enumerate}
        \item Dealing with wrong data types (e.g. text instead of integer) by removing it altogether.
        \item Dealing with missing values, keep them as is (empty), as long as most of the log itself still contains proper values.
        \item Normalizing numeric values that were out of the range expected for certain fields, taking the average from the same classification group if possible.
        \item Correcting errors in parsing in case of incorrect syntax, etc.
    \end{enumerate}
    \item Shuffling of the logs in order to avoid any existing patterns in the way they will be split to sets.
    \item We split the data set according to the "Holdout method" to a training set, which also acts as a validation set, and a testing set, ahead of time. The dataset we use, whose dimensions are presented in Table \ref{tbl:dataset_labels}, consists of the instances of the test set (set aside initially and consisting of an equal number of malicious and benign instances) and a subset of the remaining instances of a larger collection of instances from 2019. The proportion between malicious and benign training set instances was selected so as to optimize the AUC of our cross-validation tests on the training set. The best cross validation results were obtained when the proportion of malicious instances in the training set is approximately 70\%, as shown by Table \ref{tbl:split_sets}. 
    \begin{table}[H]
            \centering
            \caption{Pre-Processing and Model Generation: Split to sets}
            \label{tbl:split_sets}
            
            \begin{tabular}{ |c|c|c|c| } 
                \hline
                Set & Malicious & Benign & Total \\
                 \hline
                 Training	& 1,336,432 & 526,725 & 1,863,157\\
                 Test (Holdout) & 301,213 & 301,213 & 602,426 \\
                \hline
            \end{tabular}
        \end{table}
    \item Prepared the high-performance training server, placing the relevant sets of documents ready to be input to the next stage in the process.
\end{enumerate}

\subsection{Feature Engineering}
\label{subsec:feature_engineering}
The main strength of our model lays in its features chosen carefully from the logs as each and every one of them contributes in the detection process.\\\\
We chose approximately 40 features, consisting of metadata fields (\ref{subsec:features_metadata}), runtime-related features (\ref{subsec:features_runtime}), and PE-related fields (\ref{subsec:features_PE}) (see sub-section \ref{sec:log_struct}). Anonymized features (\ref{subsec:features_anon}) were not selected. The full list of features is presented in \ref{ChapterAppendixC}.

\subsection{Textual Features to Associations}
After extracting the features, we wanted to understand what is the similarity between all their values in the set that we received. As the features have textual representation we decided to leverage this and use a well-known method from the field of Natural Language Processing, called \textit{word2vec}.\\\\
This method uses a shallow two-layered neural network to process and convert every value ('word') in the log to a meaningful one-dimensional vector of size 32. Using \textit{word2vec}, values ('words') with similar context should have a similar vector as given by measuring cosine similarity between values.\\
We wanted a simple pre-tweaked black box solution for the word embedding, so we tested both of the popular options: \textit{word2vec} \cite{wikipedia_word2vec} and \textit{FastText}\cite{fasttext}, there were only minor differences between them in terms of the resulting detection quality, so we chose to use \textit{word2vec}.\\\\

\begin{figure}[H]
        	\begin{center}
        		\includegraphics[width=0.55\columnwidth]{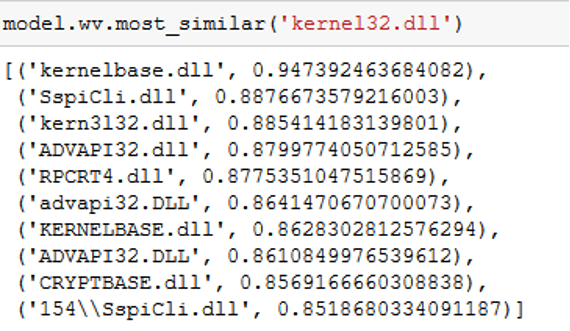}
        	\end{center}
        	\caption{Pre-Processing and Model Generation: Vector value for "kernel32.dll"}
            \label{figure:kernel32dll}
\end{figure}

In Figure \ref{figure:kernel32dll}, we can see an example of the most similar vectors for the vector representing the string "kernel32.dll" from the feature "list of loaded modules".

\begin{figure}[H]
        	\begin{center}
        		\includegraphics[width=0.7\columnwidth]{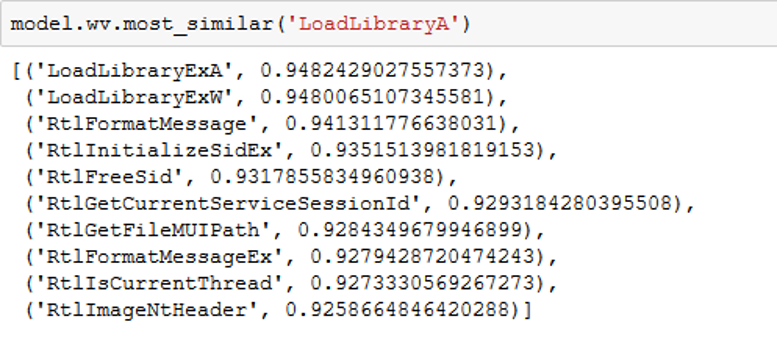}
        	\end{center}
        	\caption{Pre-Processing and Model Generation: Vector value for "LoadLibraryA"}
        	\label{figure:loadlibrarya}
\end{figure}

In Figure \ref{figure:loadlibrarya}, we can see an example of the most similar vectors for the vector representing the string "LoadLibraryA" which is a common Win32 API function used by both malware and benign applications.

The \textit{word2vec} model is trained in an unsupervised manner, which means we ignore the labels (benign, malicious), and provide a vector for each and every unique feature value in the training set.

\subsection{Dimensionality Reduction}
After the \textit{word2vec} model is trained, we need to convert every log as a whole to a representative vector that we can work with. For this purpose we use a common method usually used for document classification tasks, which allows us to reduce the dimension of the log while preserving semantics.\\\\
For each of the following types of features, we calculated the mean vector of all the vectors (that we received as an output from the \textit{word2vec} model) that represents the values ('words') inside them using Python scripting: 
\begin{enumerate}
    \item Stack snapshot and trace
    \item Registers data, opcodes collected in memory
    \item Loaded modules and resources
    \item Process and metadata information (e.g. parent process and OS version, respectively)
\end{enumerate}

This produces a one-dimensional vector of size 192 for every specific log. This vector is then used as input for a supervised classifier to be trained in the next stage.
\begin{figure}[H]
        	\begin{center}
        		\includegraphics[width=0.7\columnwidth]{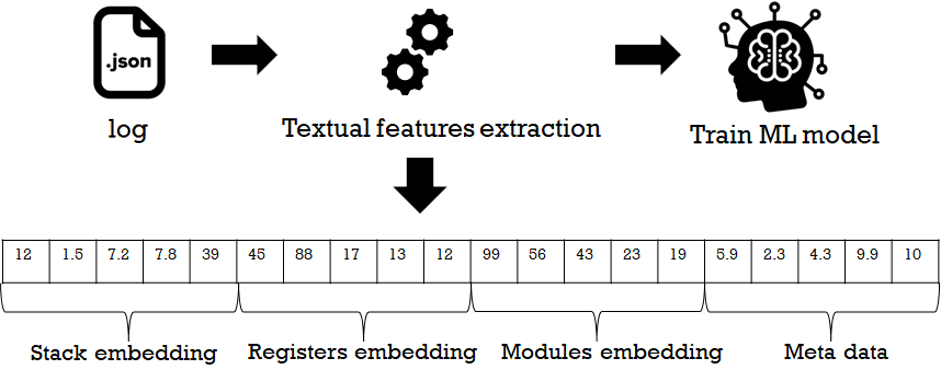}
        	\end{center}
        	\caption{Pre-Processing and Model Generation: \textit{log2vec}}
        	\label{figure:vector_log}
\end{figure}

Figure \ref{figure:vector_log}\space\space illustrates the process of transforming a single log to a one-dimensional vector using the vectors we got from \textit{word2vec}, as well as an example of how this vector is organised.

\subsection{Training}
Now that we have all the logs represented as one-dimensional vectors in the training set, we can input them as training data with their labels, to an ML algorithm trained in a supervised manner.\\
We chose the gradient boosting technique which outputs a prediction model in the form of an ensemble of weak prediction models (in our case, decision trees \cite{wikipedia_decision_tree}).\\
We wanted a simple pre-tweaked black box solution for gradient boosting, so we tested both of the popular options: \textit{XGBoost} \cite{wikipedia_xgboost} and \textit{LightGBM} \cite{wikipedia_lightgbm}. Since there were only minor differences between them in terms of detection quality, we chose \textit{LightGBM} which was easier to use.

\section{Detection Framework}
\label{SectionSolution}
\subsection{Architecture}
Our detection framework consists of the following parts:\\\\
\textbf{I. "Detector"}: A novel machine learning model trained on a big data set of environmental run-time logs captured from real-world malicious incidents per a specific protected process. This model runs on a standard server in the cloud, receives a log for detection and replies with a relative score, quantifying the probability that the activity represented by the log is malicious.\\\\
\textbf{II. "Agent"}: A lightweight service running on the endpoint machine whose task is to generate log instances from critical processes that are to be protected and send them to the Detector over the network.\\
\subsection{High level flow}
\begin{enumerate}
    \item A critical process, which should be protected, is running on the endpoint machine, a lightweight log extraction service is installed and extracts logs during run-time as required.
    \item The extracted log is sent from the endpoint machine to the Detector server for detection.
    \item The Detector server inputs the received log into the already-deployed machine learning model to receive a detection score.
    \item A result with a score for the activity diagnosed from the given log, is being sent from the server to the endpoint for taking further actions as needed.
\end{enumerate}

\begin{figure}[H]
\begin{center}
\includegraphics[width=0.8\columnwidth]{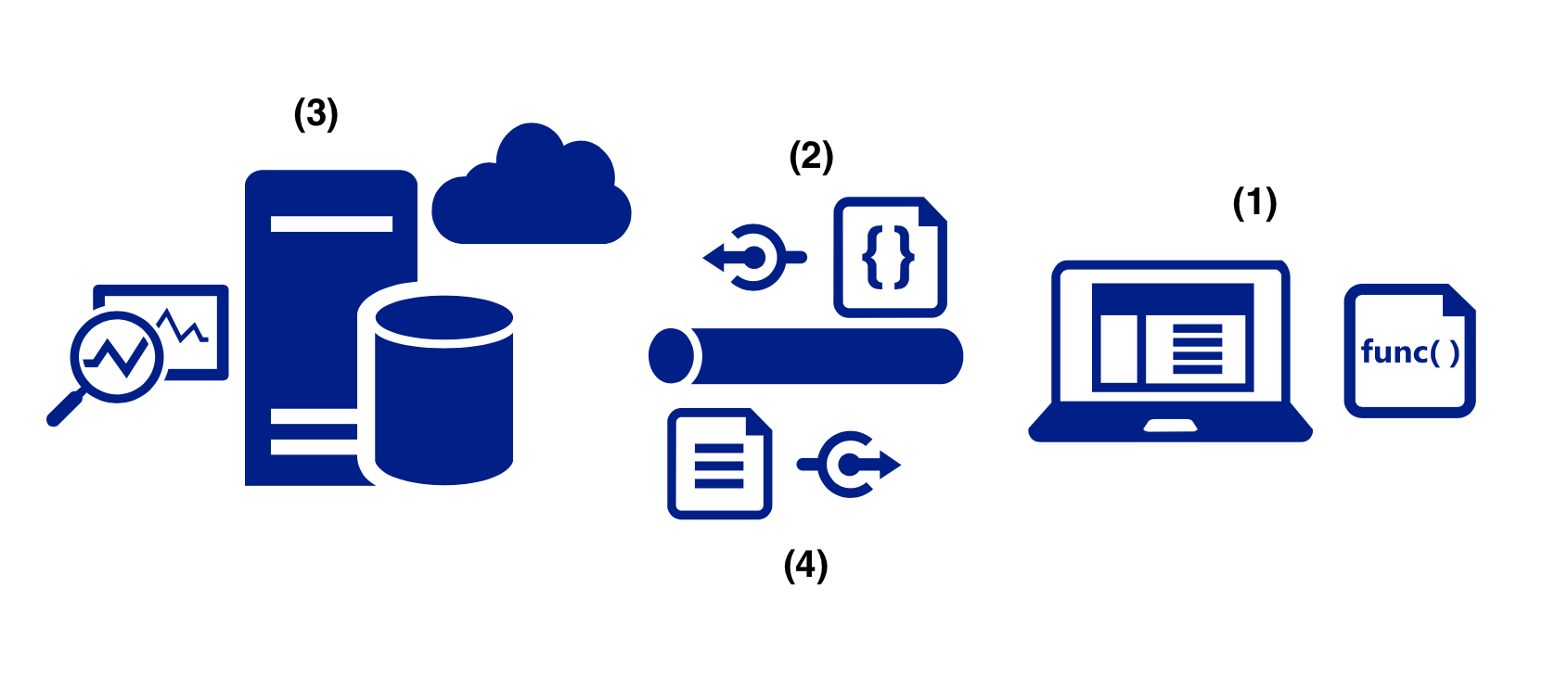}
\caption{Detection Framework: High Level Flow.}
\label{figure:hl_sol}
\end{center}
\end{figure}

\section{Experimental Evaluation}
\label{SectionExperimentalEvaluation}
In this section, we describe the experimental evaluation we conducted to assess the quality of our detection framework and its results.
\subsection{Experiment 1: Detection Model Confusion Matrix}
\begin{figure}[H]
        	\begin{center}
        		\includegraphics[width=0.55\columnwidth]{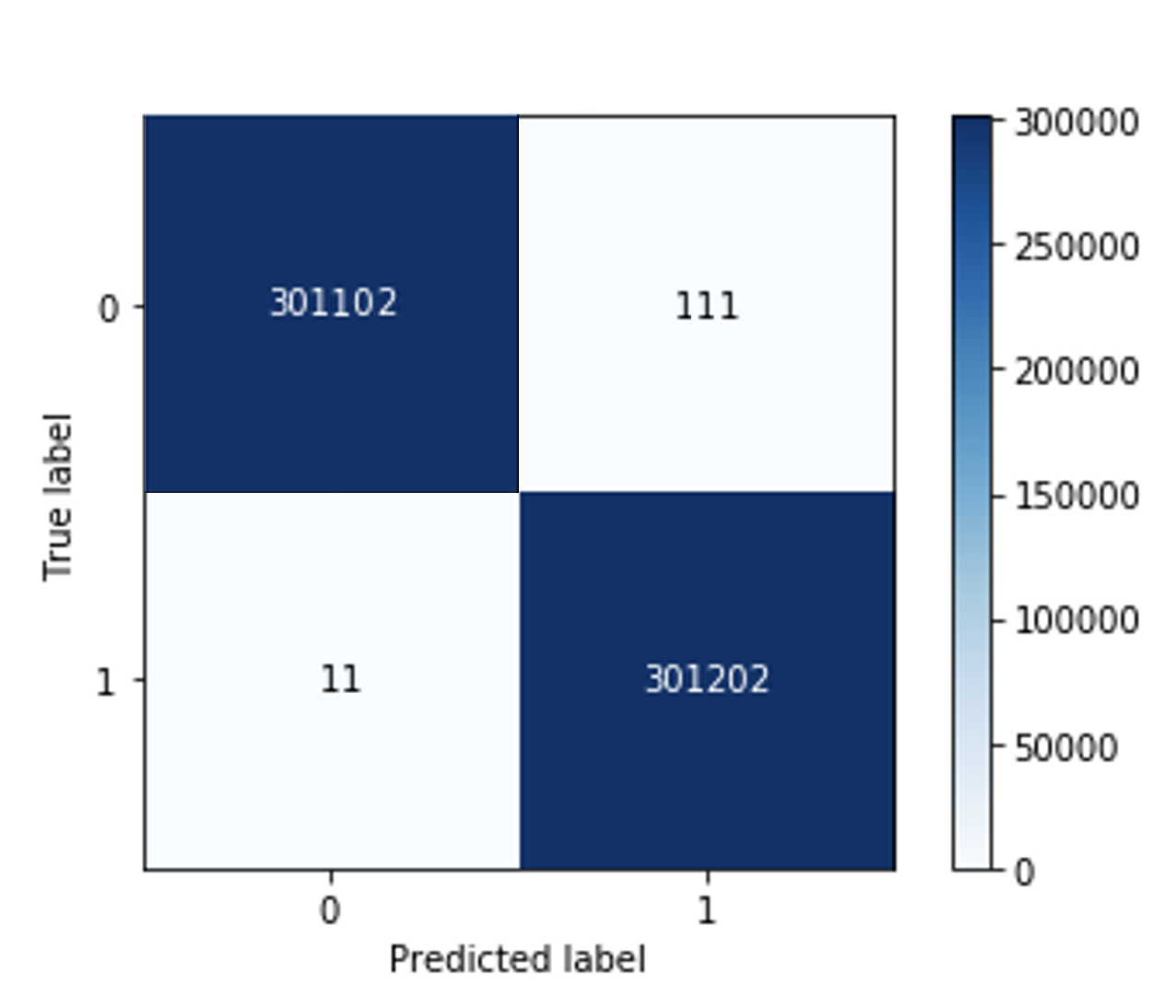}
        		\caption{Experimental Evaluation: Confusion heatmap}
        		\label{figure:test_heatmap}
        	\end{center}
\end{figure}

Our evaluation uses the test set defined in section \ref{SectionMethod}. The results are presented by Table \ref{tbl:test_results} and Figure \ref{figure:test_heatmap}. These results were obtained by using a classification threshold of 0.75.
\begin{table}[H]
            \centering
            \caption{Experimental Evaluation: Confusion Matrix for the test set}
            \label{tbl:test_results}
            \small
            \begin{tabular}{ |c|c|c| } 
                \hline
                 & Positive (predicted) & Negative (predicted) \\
                \hline
                Positive (actual) & 301,202 & 11 \\
                \hline
                Negative (actual) & 111 & 301,102 \\
                \hline
            \end{tabular}
        \end{table}
Other metrics that are computed based on the confusion matrix are shown in Table \ref{tbl:test_metrics}.
\begin{table}[H]
            \centering
            \caption{Experimental Evaluation: Metrics for the test set}
            \label{tbl:test_metrics}
            
            \begin{tabular}{ |c|c| } 
                \hline
                 Metric & Value \\
                \hline
                Area Under the ROC Curve (AUC) & 0.997712 \\
                Accuracy (ACC) & 0.999797 \\
                Precision (PPV) & 0.999632 \\
                Recall (TPR) & 0.999963 \\
                False Positive Rate (FPR) & 0.000369 \\
                False Negative Rate (FNR) & 0.000037 \\
                F1 Score & 0.999798 \\
                \hline
            \end{tabular}
        \end{table}

As presented by Table \ref{tbl:test_metrics}, detection quality is very high. Specifically, the recall is nearly 1 while the FPR is extremely low, approximately 0.04\%.

\subsection{Experiment 2: Testing Benign Applications}
\label{subsec:experimentbenign}
\begin{figure}[H]
        	\begin{center}
        		\includegraphics[width=0.45\columnwidth]{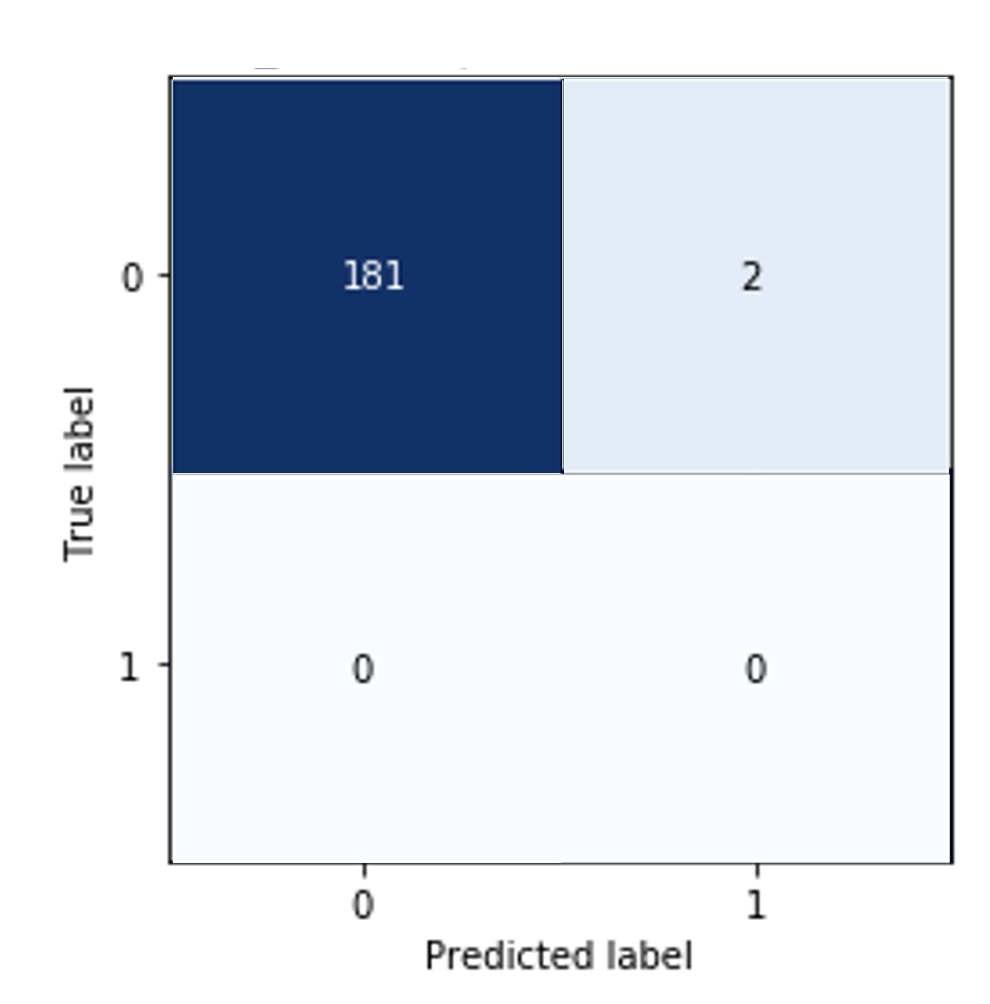}
        		\caption{Experimental Evaluation: Benign applications}
        		\label{figure:test_benign_apps}
        	\end{center}
\end{figure}

We already tested our solution using logs from the dataset that represent benign processes, however, we wanted to simulate a real-world scenario with a typical machine in an organisation, running a mixture of productivity applications, collaboration tools, etc., in parallel.\\\\
For this endeavour, we installed a pre-defined set of benign applications on a Windows 10 64-bit machine, all of them are either mainstream applications intended for daily use, or already built in the operating system.\\\\
As the list contains \textbf{30+} applications, it is included in \ref{ChapterAppendixA}.
We covered all of these applications using our Agent service, collecting logs intermittently, and assessing the results returned from the Detector in real time.\\\\
Using the same \textbf{0.75} classification threshold, we placed the \textbf{184} results in a confusion heatmap . As seen in Figure \ref{figure:test_benign_apps}, results are very good, having only \textbf{2} logs as false-positives, as \textbf{181} of the logs were classified correctly as benign.

\subsection{Experiment 3: Testing Malicious Applications}
\label{subsec:experimentmalicious}
\begin{figure}[H]
        	\begin{center}
        		\includegraphics[width=0.45\columnwidth]{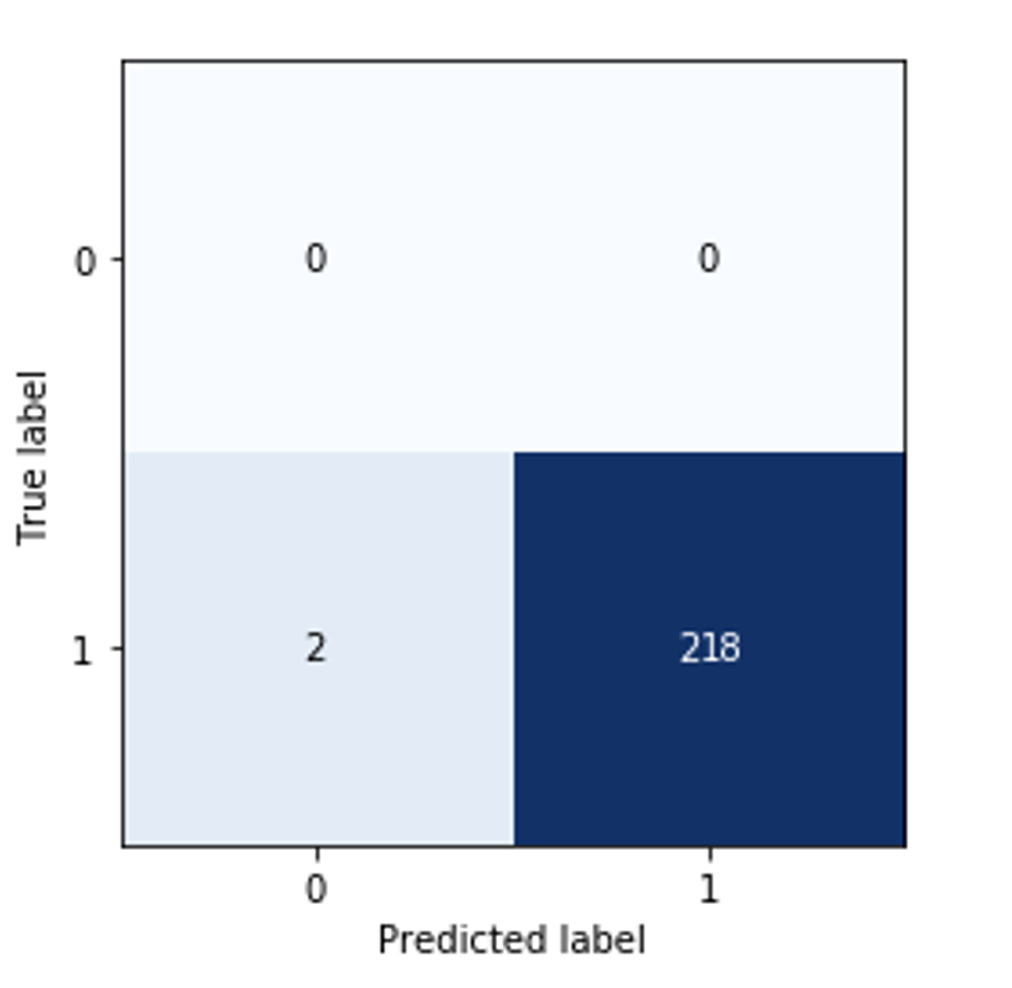}
        		\caption{Experimental Evaluation: Malware}
        		\label{figure:test_malicious_apps}
        	\end{center}
\end{figure}

In this experiment, we simulated a real-world scenario with a typical machine in an organisation, infected by malware from different types.\\\\
For this endeavour, we used an automation environment (Cuckoo sandbox \cite{cuckoo}) to run malware on a Windows 10 virtual machine, while the Agent service is also running to collect logs.\\\\
As the list contains \textbf{40+} examples of different malware, it is included in  \ref{ChapterAppendixB}, sorted by categories. We repeated the experiment for each of these malicious programs.\\\\
Using the same \textbf{0.75} classification threshold, we placed the \textbf{220} results in a confusion heatmap. As can be seen in Figure \ref{figure:test_malicious_apps}, only \textbf{2} logs were misclassified as false negatives, whereas \textbf{218} of the logs were correctly classified as malicious.

\subsection{Experiment 4: Analyzing Performance}
In order to assess how efficient our Agent is and if it causes any noticeable degradation in overall system performance, we conducted two tests:
\begin{enumerate}
    \item Running standard applications typically used by enterprise users, with and without the Agent coverage for these applications, for a few hours, while collecting logs intermittently.
    \item Calculating the amount of CPU and memory usage for the Agent service while continuously collecting logs.
\end{enumerate}
The list of applications tested can be viewed in Table \ref{tbl:test_apps_list}, the resulting performance graphs appear in \ref{ChapterAppendixD}.
\begin{table}[H]
            \centering
            \caption{Experimental Evaluation: Applications tested for performance}
            \label{tbl:test_apps_list}
            \small
            \begin{tabular}{ |c c| } 
                \hline
                 Application & \\
                \hline
                Microsoft Excel 2016 & Acrobat Reader DC \\
                Microsoft PowerPoint 2016 & Skype (8.55) \\
                Microsoft Word 2016 & Internet Explorer 11 \\
                Microsoft Outlook 2016 & Google Chrome (v78) \\
                \hline
            \end{tabular}
        \end{table}
According to our testing, if we ignore the outliers, there was no noticeable overhead running the applications in test with the agent compared to running them without the Agent.\\\\
As the Agent's CPU utilization is always below 100\%, it is using one logical CPU core at most. In addition, the Agent's memory usage never exceeded 25MB, which is reasonably low on contemporary commodity computers.

\section{Conclusion}
\label{SectionConclusionFuture}
As malware rose in sophistication and volume in recent years, so is the requirement for innovative solutions for its detection. In this work we presented an end to end ML-based solution for the early detection of in-memory malicious activity via runtime logs, to demonstrate the potential of leveraging the plethora of process and environment data that can be collected from processes executing on endpoint machines over time and be used for malware detection. Our experimental evaluation establishes that this approach has the potential of yielding high-quality detection of malicious process activity before it performs harmful operations.\\
In future work, we plan to investigate the length of the early warning period that can be achieved using our detection approach, that is, how long before the malware attempts to execute its harmful operations can it be detected using our approach. In addition, there are challenges yet to be overcome regarding our solution's limitations, these include: reducing the downtime on update, minimizing compatibility issues on the Agent's side, attempting to provide offline detections, overcoming the performance limitation of parallel logging and the overhead stemming from retraining of the model. Additional avenues for future research include extending our detector by improving the server-side component, security, communications, anti tampering, and logging mechanisms.

\begin{appendices}
\renewcommand{\thesection}{\appendixname~\Alph{section}}

\section{Logs}
\label{ChapterAppendixC}
In this appendix, we list the log fields used as features in sub-section \ref{subsec:feature_engineering}.
\subsection{Anonymized}
\label{subsec:features_anon}
The following fields contain personal information and are anonymized in advance:
\begin{enumerate}
    \item \textbf{Username} of the user logged in (if there is an interactive user logged in) at the time of the incident.
    \item \textbf{Domain name} the target machine belongs to (if there is one).
    \item \textbf{Machine name} of the target machine.
    \item \textbf{IP address} of the target machine.
    \item \textbf{Machine serial number} if the machine is prebuilt by a vendor or an OEM (e.g. a laptop).
\end{enumerate}

\subsection{Metadata}
\label{subsec:features_metadata}
The following fields are considered metadata:
\begin{enumerate}
    \item \textbf{Timestamp} of the incident, Date/Time in milliseconds.
    \item \textbf{Operating System} of the target machine, in our case, a version of Windows (e.g. Windows 7 32-bit).
    \item \textbf{Operating System build number}
    \item \textbf{Executable full path} for the executable whose code the process under observation is executing.
    \item \textbf{Executable name}.
    \item \textbf{Executable hash}.
    \item \textbf{File attributes} Creation time and modification time of the executable.
    \item \textbf{Referral URL} from where the executable was downloaded, if it was downloaded off the internet.
    \item \textbf{User login time} of the last logged in user.
    \item \textbf{Thread count} for running threads in the process.
    \item \textbf{Integrity level} of the process, can be Untrusted, Low, Medium, High or System. This defines the trust level for this process in Windows based systems.
    \item \textbf{Executable architecture} whether it is 32 or 64 bit.
    \item \textbf{Total work time} done for the process so far in both processor clock cycles and in process time done in kernel mode.
    \item \textbf{Process ID} of the executable.
    \item \textbf{Thread ID} of the thread executing the monitored code within the process.
    \item \textbf{User privilege level} for the currently logged in user, if there is one, allows distinguishing between Standard users, Guests and Administrators.
    \item \textbf{Timezone} of the target machine at the time of the incident.
\end{enumerate}

\subsection{Runtime}
\label{subsec:features_runtime}
The following fields are considered essential information for understanding the incident and are collected from the environment and the process itself during runtime:
\begin{enumerate}
    \item \textbf{Base address} starting from which the executable was mapped into memory.
    \item \textbf{Command line arguments} used for running the executable.
    \item \textbf{Value of all the registers}, that is, a the current contents of EAX, EBX, ECX, EDX, EBP, ESP, EDI, ESI, etc. for the currently executing thread in the process.
    \item \textbf{Snippets of bytes pointed to by all the registers}, that is, a snapshot of the bytes pointed by the addresses in registers EAX, EBX, ECX, EDX, etc. if these are valid addresses in memory, for the current executing thread (although the heap, code and data are shared across the process).
    \item \textbf{EFLAGS register contents} for the currently executing thread in the process.
    \item \textbf{Digital certificate / signature} of the executable.
    \item \textbf{List of loaded resources} by the executable, and their properties: full path, size, hash, creation time, modification time.
    \item \textbf{Amount of virtual memory} both free and used in bits.
    \item \textbf{Specific information from HKLM registry values} including "run" and "runonce" key-value pairs.
    \item \textbf{DEP protection} status at the time of the incident.
    \item \textbf{List of illegal addresses in the virtual memory} the executable tried to access and snippets of bytes from these addresses, hopefully translating into opcodes.
    \item \textbf{Hash of the import table} from the PE file.
    \item \textbf{Details about the injecting process}, if there is one (e.g. remote thread injection), details include: process ID, parent process ID, hash, full path to executable.
    \item \textbf{Flag indicating if the executable has the auto-elevation} property in its manifest. This property is typically used for running at a higher privilege level in order to bypass the User Account Control on Windows systems.
    \item \textbf{List of loaded modules} at the time of the incident, including details for each one: base address in memory, end address in memory, size, file linking metadata and full path to the module.
    \item \textbf{List of opened resources} by the process in the time of the incident, for example, file handles, and the attributes of these files.
    \item \textbf{Details about the parent process} of the executable, including process ID, full path, hash, file attributes, command line arguments used to run the parent process, integrity level, resources with their details, loaded modules and opened resources.
    \item \textbf{List of process block data} from the executable up to the first unprotected parent process.
    \item \textbf{Snapshot of the stack's memory and stack trace} at the time of the incident for the current executing thread.
    \item \textbf{List of files by their magic} if found in the process's memory, e.g. PDF or script files in the memory of the process.
    \item \textbf{List of URLs}, if found, using a regex search in the process' memory.
    \item \textbf{List of IP addresses}, if found, using a regex search in the process' memory.
    \item \textbf{List of task scheduler tasks} if they are related to the executing process.
    \item \textbf{List of registry change attempts} and their results in certain regions of the registry.
\end{enumerate}

\subsection{PE}
\label{subsec:features_PE}
The following fields are extracted from the PE file itself, if possible:
\begin{enumerate}
    \item \textbf{PE type}
    \item \textbf{Section count}
    \item \textbf{Section information}
    \item \textbf{Import count}
    \item \textbf{Import information}
    \item \textbf{Export count}
    \item \textbf{Export information}
    \item \textbf{Executable export name}
    \item \textbf{Characteristics}
    \item \textbf{Compilation time}
    \item \textbf{Executable signature}
    \item \textbf{Executable architecture}
    \item \textbf{Entry point}
    \item \textbf{Entropy level}
    \item \textbf{Executable size}
    \item \textbf{PDB path}
    \item \textbf{Executable Creation time}
    \item \textbf{Executable Modification time}
\end{enumerate}

\subsection{Example}
\begin{figure}[H]
        	\begin{center}
        		\includegraphics[width=0.65\columnwidth]{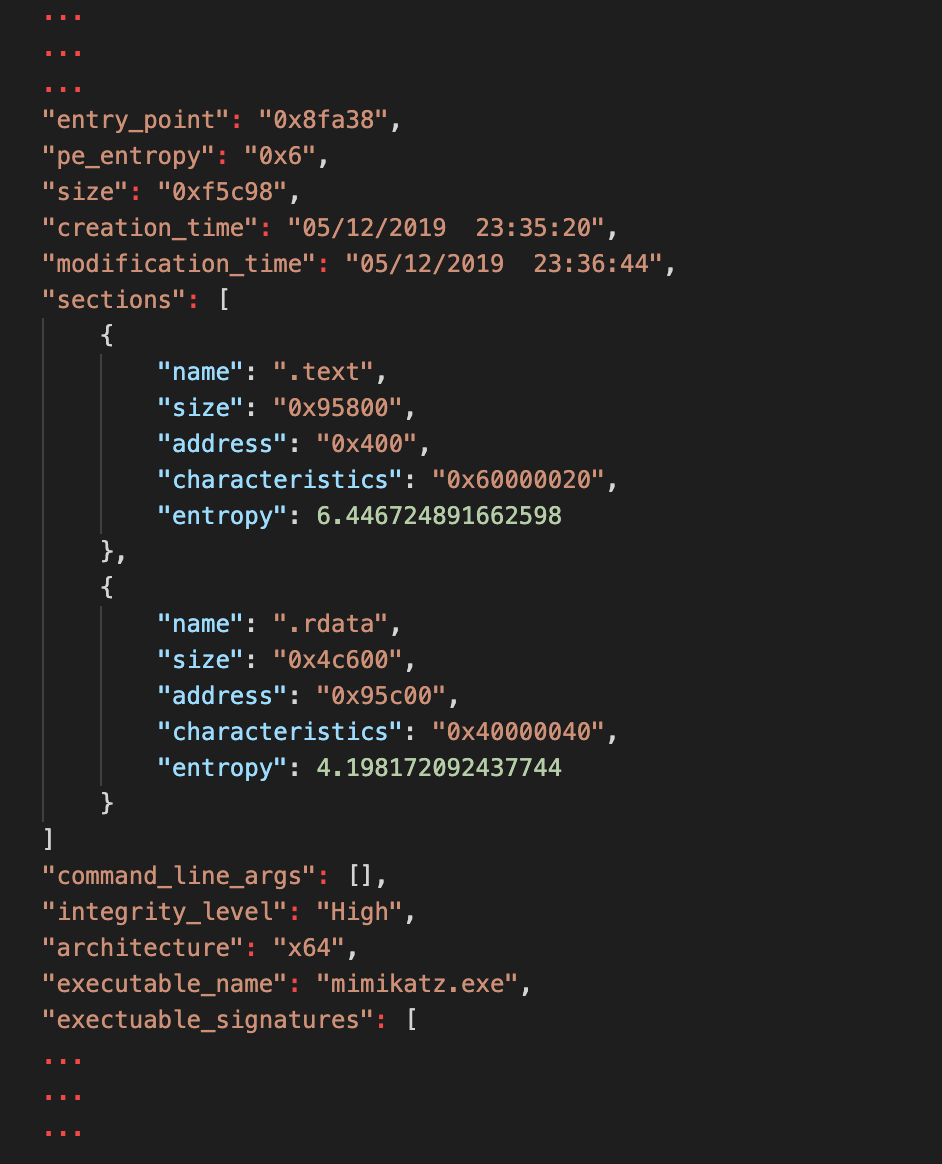}
        	\end{center}
        	\caption{\ref{ChapterAppendixC}: Log example}
        	\label{figure:log_example}
\end{figure}

In Figure \ref{figure:log_example} we can see an example of a partial log file collected in real time for a \textit{Mimikatz} credential theft attack, which is considered an info stealing malware\cite{mimikatz}.

\section{Benign Applications List}
\label{ChapterAppendixA}
In this appendix, we list the benign applications being used for testing in sub-section \ref{subsec:experimentbenign}, by category.
\begin{enumerate}
    \item Collaboration applications are listed in Table \ref{tbl:test_apps_colab}, \textbf{7} applications in total.
    \item Productivity and media applications are listed in Table \ref{tbl:test_apps_prod}, \textbf{16} applications in total.
    \item Applications used for executing code are listed in Table \ref{tbl:test_apps_code}, \textbf{10} applications in total.
\end{enumerate}

\begin{table}[H]
            \centering
            \caption{\ref{ChapterAppendixA}: Collaboration applications}
            \label{tbl:test_apps_colab}
            \small
            \begin{tabular}{ |c|c|c| } 
                \hline
                 Application & Usage & Executable Name \\
                \hline
                Cisco WebEx & Virtual conferencing & webex.exe \\
                GoToMeeting & Virtual conferencing & gotomeeting.exe \\
                Skype & Voice-over-IP & skype.exe \\
                Slack & Instant messaging & slack.exe \\
                Microsoft Teams & Virtual conferencing & teams.exe \\
                TeamViewer & Screen sharing & teamviewer.exe \\
                Zoom & Virtual conferencing & zoom.exe \\
                \hline
            \end{tabular}
        \end{table}
        
        \begin{table}[H]
            \centering
            \caption{\ref{ChapterAppendixA}: Productivity and media applications}
            \label{tbl:test_apps_prod}
            \small
            \begin{tabular}{ |c|c|c| } 
                \hline
                 Application & Usage & Executable Name \\
                \hline
                Acrobat Reader & PDF viewer & acrord32.exe \\
                Microsoft Excel & Spreadsheet editor & excel.exe \\
                Microsoft Word & Document editor & word.exe \\
                Microsoft PowerPoint & Presentation editor & powerpnt.exe \\
                Microsoft Access & Database management & msaccess.exe \\
                Microsoft Outlook & Email client & outlook.exe \\
                Google Chrome & Web browser & chrome.exe \\
                Mozilla Firefox & Web browser & firefox.exe \\
                Microsoft Internet Explorer & Web browser & iexplore.exe \\
                Microsoft Edge & Web browser & msedge.exe \\
                VideoLAN (VLC) & Media player & vlc.exe \\
                7-Zip & File archiver & 7z.exe \\
                WinRAR & File archiver & winrar.exe \\
                FortiClient & VPN & forticlient.exe \\
                Task Manager & Scheduled tasks & taskeng.exe \\
                Registry Editor & Windows Registry & regsvr32.exe \\
                \hline
            \end{tabular}
        \end{table}
        
        \begin{table}[H]
            \centering
            \caption{\ref{ChapterAppendixA}: Code-execution applications}
            \label{tbl:test_apps_code}
            \small
            \begin{tabular}{ |c|c|c| } 
                \hline
                 Application & Usage & Executable Name \\
                \hline
                AutoIt & AutoIt V3 scripts & autoit3.exe \\
                Java RTE & Java code & java.exe \\
                Python 3 & Python code & python3.exe \\
                NodeJS & NodeJS applications & node.exe \\
                Cscript & VBscript / WSH script & cscript.exe \\
                MSHTA & Microsft HTML applications & mshta.exe \\
                Rundll32 & Load DLLs & rundll32.exe \\
                Wscript & Windows Script Host & wscript.exe \\
                PowerShell & PowerShell scripts & powershell.exe \\
                CMD & Command prompt & cmd.exe \\
                \hline
            \end{tabular}
        \end{table}

\section{Malicious Applications List}
\label{ChapterAppendixB}

In this appendix, we list the malware being used for testing in sub-section  \ref{subsec:experimentmalicious}, by category, using either their common name or their CVE.
\begin{enumerate}
    \item Malware exploiting the target system via PDF files, opened via Acrobat Reader, are listed in Table \ref{tbl:test_mal_pdf}, \textbf{8} attacks in total.
    \item Malware exploiting the target system via docx or xlsx files, opened via Microsoft Word or Excel, are listed in Table \ref{tbl:test_mal_docx}, \textbf{15} attacks in total.
    \item Malware exploiting the target system via VB, Flash or Silverlight, are listed in Table \ref{tbl:test_mal_vb}, \textbf{30} attacks in total.
    \item Exploit kits are listed in Table \ref{tbl:test_mal_ek}, \textbf{8} exploit kits in total.
    \item Ransomware are listed in Table \ref{tbl:test_mal_ransomware}, \textbf{43} attacks in total.
    \item Various other malware are listed in Table \ref{tbl:test_mal_va}, \textbf{56} attacks in total.
\end{enumerate}

\begin{table}[H]
            \centering
            \caption{\ref{ChapterAppendixB}: PDF malware}
            \label{tbl:test_mal_pdf}
            \small
            \begin{tabular}{ |c c c| } 
                \hline
                 CVE & & \\
                \hline
                CVE-2014-0496 & CVE-2013-0640 & CVE-2010-2883 \\
                CVE-2013-3346 & CVE-2013-2729 & CVE-2010-0188 \\
                CVE-2012-0754 & CVE-2011-2462 & \\
                \hline
            \end{tabular}
        \end{table}
        
        \begin{table}[H]
            \centering
            \caption{\ref{ChapterAppendixB}: docx/xlsx malware}
            \label{tbl:test_mal_docx}
            \small
            \begin{tabular}{ |c c c c| } 
                \hline
                 CVE & & & \\
                \hline
                CVE-2017-11882 & CVE-2017-11826 & CVE-2017-0262 & CVE-2015-5119 \\
                CVE-2015-2545 & CVE-2015-1770 & CVE-2014-1761 & CVE-2013-3906 \\
                CVE-2012-0158 & CVE-2012-2539 & CVE-2012-0158 & CVE-2012-1642 \\
                CVE-2012-0158 & CVE-2009-3129 & CVE-2010-3333 & \\
                \hline
            \end{tabular}
        \end{table}
        
        \begin{table}[H]
            \centering
            \caption{\ref{ChapterAppendixB}: VB/Flash/Silverlight malware}
            \label{tbl:test_mal_vb}
            \small
            \begin{tabular}{ |c c c| } 
                \hline
                 CVE & & \\
                \hline
                CVE-2018-15982 & CVE-2018-8174 & CVE-2018-4878 \\
                CVE-2017-11292 & CVE-2015-7645 & CVE-2015-5560 \\
                CVE-2015-5122 & CVE-2015-5119 & CVE-2015-3113 \\
                CVE-2015-3105 & CVE-2015-3104 & CVE-2015-3090 \\
                CVE-2015-0359 & CVE-2015-0336 & CVE-2015-0313 \\
                CVE-2015-0311 & CVE-2014-8439 & CVE-2014-0569 \\
                CVE-2014-0556 & CVE-2014-0515 & CVE-2014-0497 \\
                CVE-2013-3897 & CVE-2013-3893 & CVE-2013-3163 \\
                CVE-2013-2551 & CVE-2013-1347 & CVE-2012-4792 \\
                CVE-2012-1876 & CVE-2012-1875 & CVE-2011-1996 \\
                \hline
            \end{tabular}
        \end{table}
        
        \begin{table}[H]
            \centering
            \caption{\ref{ChapterAppendixB}: Exploit kits}
            \label{tbl:test_mal_ek}
            \small
            \begin{tabular}{ |c c| } 
                \hline
                 Name & \\
                \hline
                Angler & Nuclear \\
                Fiesta & RIG \\
                Magnitude & Sundown \\
                Neutrino & Sweet Orange\\
                \hline
            \end{tabular}
        \end{table}
        
        %\begingroup
        \begin{table}[H]
        \centering
        \caption{\ref{ChapterAppendixB}: Ransomware}
        \label{tbl:test_mal_ransomware}
        \small
        \begin{tabular}{|c c c c c|}
            \hline
            Name & & & & \\
            \hline
            Ryuk & Ransomware-X & RagnarLocker & Snake (Go) & Thanos \\
            MedusaLocker & PonyFinal (Java) & REvil & Bandarchor & BTCware \\
            Cerber & NewCerber & Cobra-Crysis & CoinMiner & CryptFile2 \\
            CryptoMix & CryptoLocker & CryptoWall & CryptoShield & CryptXXX \\
            Gandcrab & Dharma-AUDIT & GlobeImposter & Gryphon & Jaff \\
            JagerDecryptor & Locky & Osiris & Magniber & Matrix \\
            Mole & Petya, Mischa & Philadelphia & Santa & Satan \\
            SAVEFiles & Sega2.0 & Shade & Sigma & Spora \\
            TeslaCrypt & WannaCry & XData &  & \\
            \hline
        \end{tabular}
        \end{table}
        %\endgroup
        
        % \begingroup
        \begin{table}[H]
        \centering
        \caption{\ref{ChapterAppendixB}: Various malware}
        \label{tbl:test_mal_va}
        \small
        \begin{tabular}{| c c | c c | c c |}
                \hline
                 Name & Category & Name & Category & Name & Category \\
                \hline
                Andromeda & Bot & Bunitu Proxy & trojan & Chthonic & Banking trojan \\
                AZORULT & Spyware & BetaBot & Info stealer & BitCoinMiner & Miner \\

                Corebot & Banking trojan & Cridex & Bot, Info stealer & Cutwailpushdo & Bot \\
                DanaBOT & Bot & Dreambot-Ursnif & Banking trojan & Dridex & Info stealer \\
                Emotet & Trojan & Flawed Ammyy & RAT & FlokiBot & Banking trojan \\
                Formbook & Trojan & Godzilla & Loader & Graybird & Backdoor \\
                IcedID & Banking trojan & Imminent & RAT & ISRstealer & Info stealer \\
                Kasidet-Neutrino & Bot & Kovter & Fileless & Kronos & Banking trojan \\
                LatentBot & RAT & LokiBot & Banking trojan & Neurevt & Info stealer, Bot \\
                NSIS-injector & Trojan & PlugX & RAT & Pony & Info stealer \\
                Potao Express & Toolkit & Poweliks & Fileless & Pushdo & Bot \\
                QBot & Bot & Quant & Loader & Ramnit & Banking trojan \\
                Redaman & Banking trojan & Remcos & RAT & RokRAT & RAT \\
                Rombertik & Spyware & SmokeLoader & Bot & Snatch & Loader \\
                SpyWare & Rootkit & Terdot.A & Trojan & Zloader & Loader \\
                Tesla & Keylogger & TrickBot & Trojan & Trojan—pzamd & Trojan \\
                Uptare & Trojan & Ursnif & Banking trojan & Vawtrak & Banking trojan \\
                XMRig & Miner & ZeroAccess & Trojan & Zeus Panda & Banking trojan \\
                ZeusVM & Banking trojan & & & & \\
                \hline
            \end{tabular}
            \end{table}
        % \endgroup
        
\section{Performance}
\label{ChapterAppendixD}
\begin{figure}[H]
        	\begin{center}
        		\includegraphics[width=0.6\columnwidth]{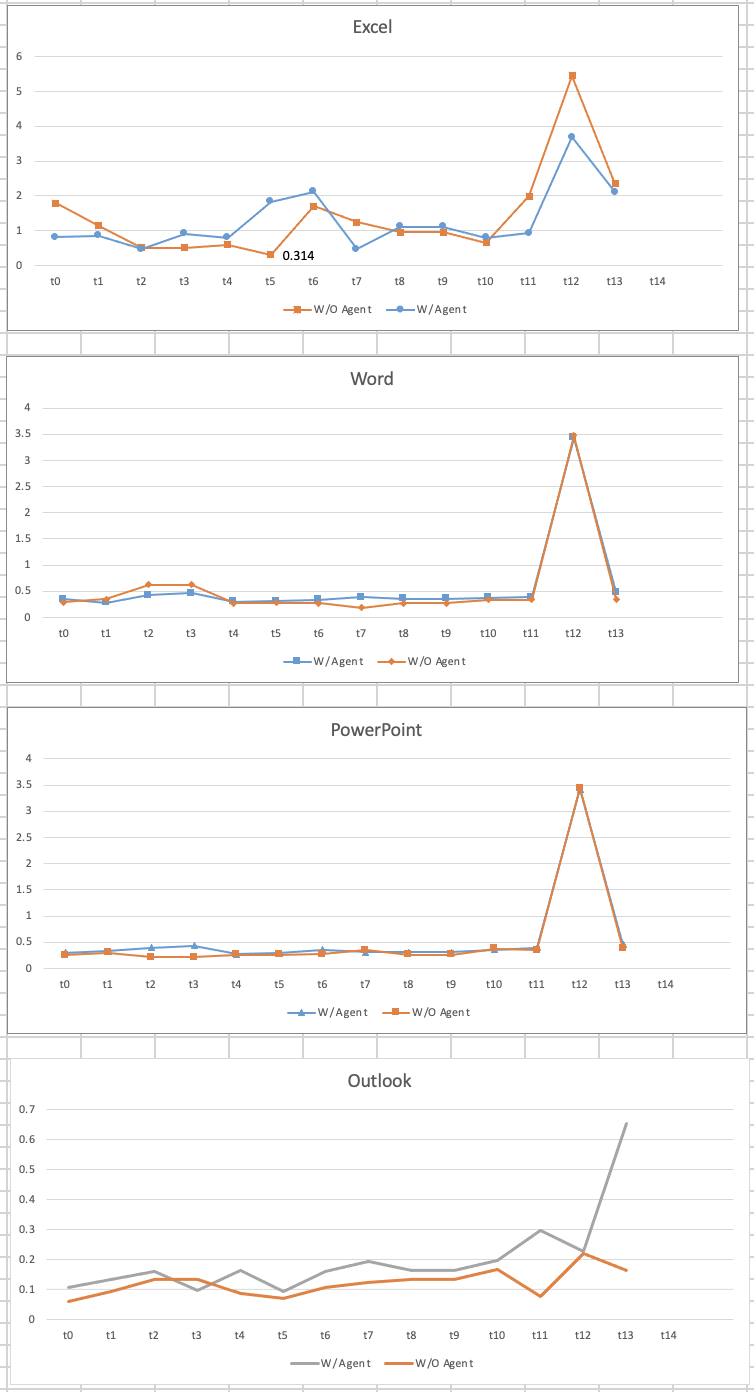}
        	\end{center}
        	\caption{\ref{ChapterAppendixD}: Microsoft Office applications with and without Agent coverage}
            \label{figure:test_apps1}
\end{figure}

\begin{figure}[H]
        	\begin{center}
        		\includegraphics[width=0.6\columnwidth]{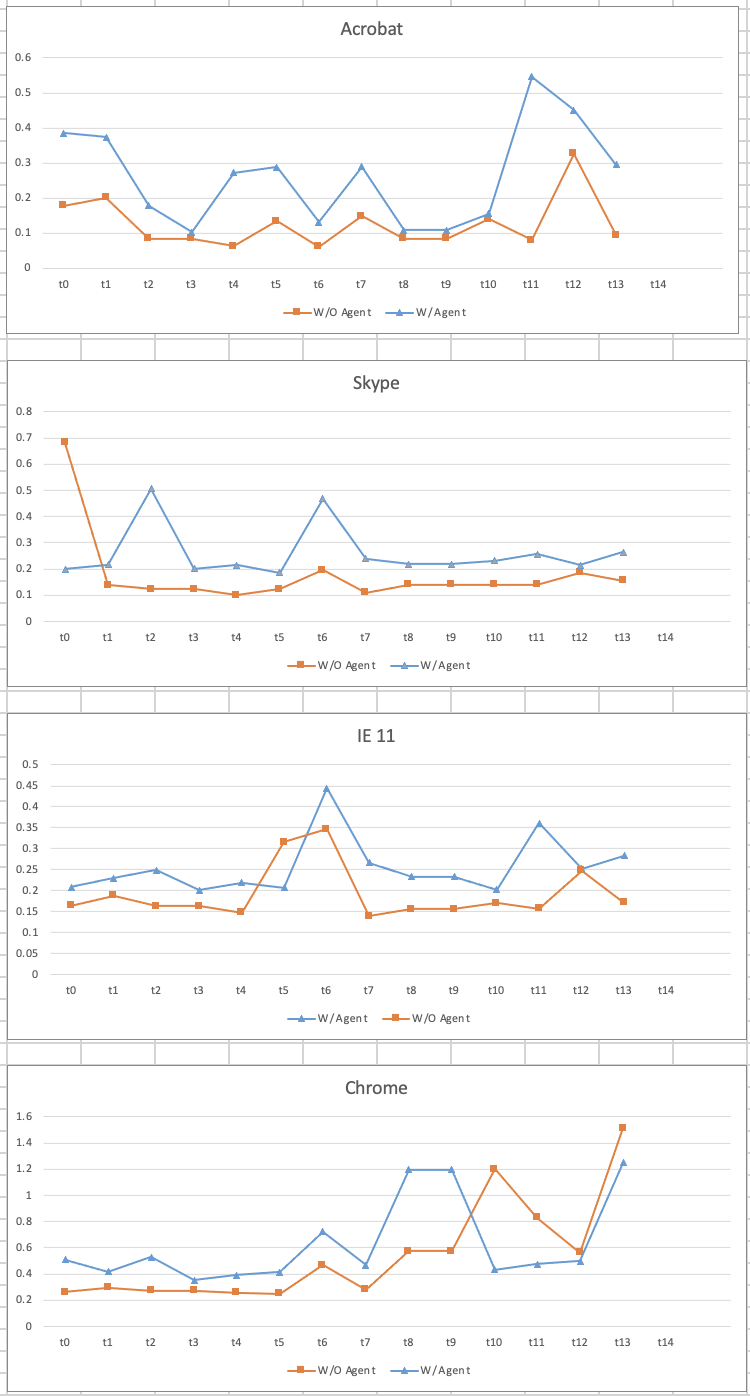}
        	\end{center}
        	\caption{\ref{ChapterAppendixD}: Various applications with and without Agent coverage}
            \label{figure:test_apps2}
\end{figure}

\begin{figure}[H]
        	\begin{center}
        		\includegraphics[width=0.6\columnwidth]{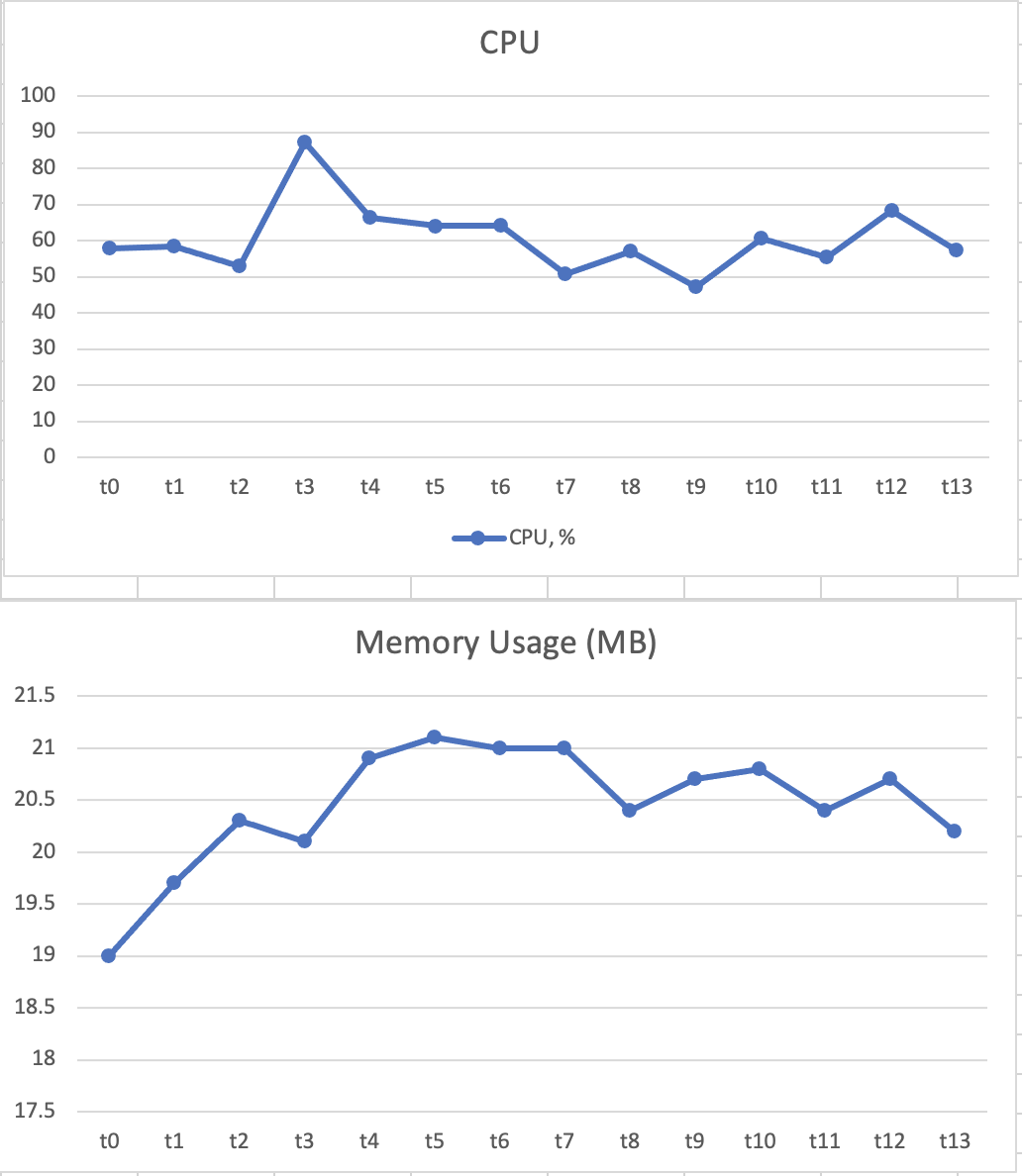}
        	\end{center}
        	\caption{\ref{ChapterAppendixD}: CPU and Memory Usage}
            \label{figure:test_log_perf}
\end{figure}

In this appendix, we provide the results of the performance analysis done for the Agent.\\ 
Figure \ref{figure:test_apps1}\space\space presents Microsoft Office applications (e.g. Word) running with and without the Agent coverage.
Figure \ref{figure:test_apps2}\space\space presents various applications (e.g. Google Chrome), running with and without the Agent coverage.
Figure \ref{figure:test_log_perf}\space\space presents the amount of CPU usage and memory usage for the Agent service while it is collecting logs in a continuous matter.
The Y-axis of the CPU chart is the percentage of CPU utilization at time \textit{t} (the X-axis).
The Y-axis of the Memory Usage chart is the peak amount of physical RAM allocated to the Agent at time \textit{t} (the X-axis).
The tests were conducted using Microsoft Windows Performance Toolkit (aka Xperf), on the same baseline machine having the following specifications:
\begin{enumerate}
    \item OS: Windows 10 x64
    \item CPU: Intel 4-Cores
    \item RAM: 8GB DDR4
    \item SSD: 256GB SATA3
\end{enumerate}

\end{appendices}

\end{document}